\documentclass[preprint,12pt]{elsarticle}

\usepackage{float}

\journal{Radiation Detection Technology and Methods}

\biboptions{sort&compress}
\begin{document}

\begin{frontmatter}

\title{The development of $^{222}$Rn detectors for JUNO prototype}

\renewcommand{\thefootnote}{\fnsymbol{footnote}}
\author{
Y.P.~Zhang$^{a,b}$,
J.C.Liu$^{a}$,
C.~Guo$^{a}\footnote{Corresponding author. Tel:~+86-01088236256. E-mail address: guocong@ihep.ac.cn (C.~Guo).}$,
Y.B.~Huang$^{a,b}$,
C.~Xu$^{c}$,
M.Y.~Guan$^{a}$,
C.G.~Yang$^{a}$,
P.~Zhang$^{a}$
}
\address{
${^a}$Key Laboratory of Particle Astrophysics, Institute of High Energy Physics, Chinese Academy of Science, Beijing, China\\
${^b}$ School of Physics, University of Chinese Academy of Science, Beijing, China \\
${^c}$ School of Nuclear Science and Technology, University of South China, Hengyang, China\\
}

\begin{abstract}
The radioactive noble gas $^{222}$Rn, which can be dissolved in water, is an important background source for JUNO. In this paper, based on the water system of JUNO prototype, two kinds of high sensitivity radon detectors have been proposed and developed. The sensitivity of Si-PIN Rn detector, which uses a Si-PIN photodiode to detect the $\alpha$ from $^{214}$Po decay, is $\sim$9.0~mBq/m$^3$. The sensitivity of LS Rn detector, which uses liquid scintillator to detect the coincident signals of $\beta$ from $^{214}$Bi decay and $\alpha$ from $^{214}$Po decay, is $\sim$64.0~mBq/m$^3$. Both of the two kinds of Rn detector have the potential to be developed as an online Rn concentration monitoring equipment for JUNO veto detector.
\end{abstract}


\end{frontmatter}

%\linenumbers

\section{Introduction}
The Jiangmen Underground Neutrino Observatory(JUNO) is a multipurpose neutrino experiment designed to determine neutrino mass hierarchy and precisely measure oscillation parameters by detecting reactor neutrinos from the Yangjiang and Taishan Nuclear Power Plants, with a 20-thousand-tons liquid scintillator(LS) detector of unprecedented 3\% energy resolution (at 1 MeV) at 700-meter deep underground. To suppress radioactivity and cosmogenic background, the outer of the central detector is filled with ultra-pure water as passive shielding for radioactivity from surrounding rock and is equipped with about 2000 microchannel plate photomultiplier tubes( MCP-PMTs, 20~inchs) to form a water Cherenkov veto detector to tag muon. There also have strict requirements of intrinsic radioactivity in water.

The natural radioactive nobel gas radon($^{222}$Rn), which is continuously generated from the decay chain of $^{238}$U series, has a half life of 3.82 days and could be potentially dissolved into water, LS, and so on. According to the MC simulation for JUNO experiment requirements, the Rn concentration in the water should be less than $0.2$~Bq/m$^3$~\cite{JUNO}. The high sensitivity Rn assay is one of the essential techniques for JUNO as well as other underground experiments~\cite{borexino,superk}.

\begin{figure}[htb]
\centering
\includegraphics[width=8cm]{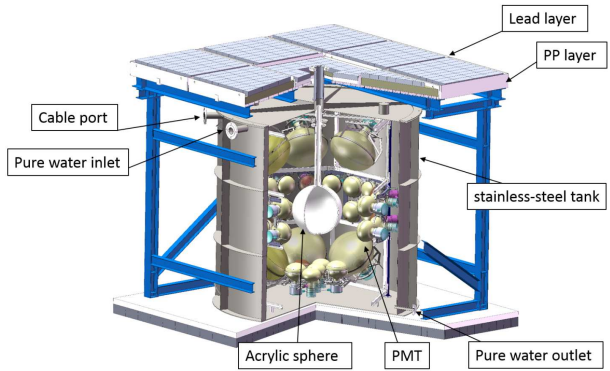}
\caption{The JUNO prototype detector. }
\label{fig.juno_prototype}
\end{figure}

\begin{figure}[htb]
\centering
\includegraphics[width=8cm]{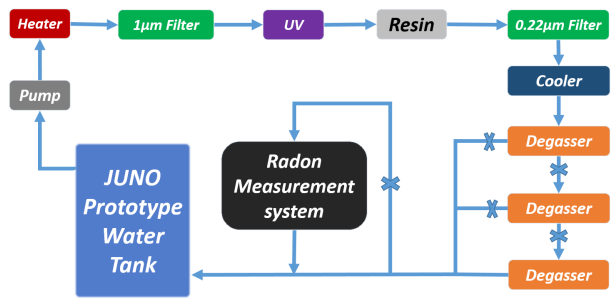}
\caption{The conceptual scheme of water system for JUNO prototype. }
\label{fig.watersystem}
\end{figure}
Considering the specific requirements of each sub-system, a prototype detector was proposed to test the key technical issues as well as to study the properties of Rn. As is shown in Fig.~\ref{fig.juno_prototype}, an acrylic sphere, which locates at the center of the stainless-steel tank (SST) as a LS vessel, is viewed by 51 PMTs which are immersed in the pure water. The PP/Lead layer is designed to have 1~meter water equivalent shielding, aiming to reduce the radioactivity coming from the outside. In order to maintain good water quality for a long time, a reliable ultra-pure water purification and circulation system is built, and the conceptual scheme is shown in Fig.~\ref{fig.watersystem}. In order to measure the $\sim0.2$~Bq/m$^3$ level Rn concentration in the water system of JUNO prototype, two kinds of high sensitivity Rn measurement systems have been developed.

\section{Experimental setup for radon measurement system}
\begin{figure}[htb]
\centering
\includegraphics[width=10cm]{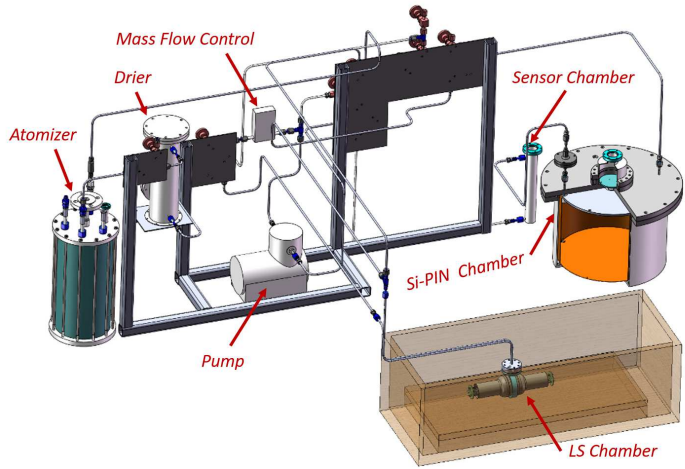}
\caption{The scheme of Rn measurement system. }
\label{fig.measurementsystem}
\end{figure}
The first step to monitor the Rn concentration in the water is to combine the measurement system and the water system. Figure~\ref{fig.measurementsystem} shows a schematic view of the combined system, which consists of the following parts:

(A) Atomizer, which is a water vapor balancing device, is used to transform the dissolved Rn gas from water into air during the flowing.

(B) Gas piping line, which consists of a drier, a mass flow control and a pump, is used to reduce the humidity of the measured gas as well as to control the gas flow rate and transfer the sample gas into the measurement devices.

(C) Monitoring system, which consists of a pressure gauge, a thermometer and a hygrometer, is used to monitor the experimental conditions.

(D) Measurement system, which consists of the Si-PIN chamber and LS chamber, is used to determine the Rn concentration.

In order to prevent the contamination of environmental air, in which the Rn concentration is around 20~Bq/m$^3$, knife-edge flanges with metal gaskets as well as VCR pipelines with metal gaskets are used to ensure the leak rate to be less than $1{\times}10^{-5}$~ml/s.

\begin{figure}[htb]
\centering
\includegraphics[width=10cm]{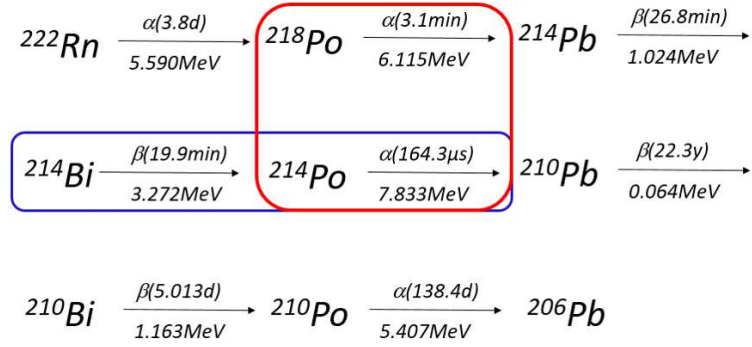}
\caption{The decay chain of $^{222}$Rn. The Si-PIN Rn detector measures the $\alpha$ from either $^{214}$Po decay or $^{218}$Po decay to determine the Rn concentration, which is shown in the red square. The LS Rn detector measure the coincident signals of $\beta$ from $^{214}$Bi decay($\beta$-$^{214}$Bi) and $\alpha$ from $^{214}$Po($\alpha$-$^{214}$Po) decay, which is shown in the blue square. }
\label{fig.decaychain}
\end{figure}

Figure~\ref{fig.decaychain} shows the decay chain of $^{222}$Rn. The principle of Si-PIN Rn detector is to collect the daughter nuclei of $^{222}$Rn to the surface of the Si-PIN in a electric field, and to measure the energy of $\alpha$s released by the collected nuclei~\cite{annreport,radioisotopes}. While for the LS Rn detector, the principle is that Rn has a great solubility coefficient in the LS and the photons produced by the $\beta$s and $\alpha$s in $^{222}$Rn decay chain can be detected by the PMTs. Furthermore, the $164.3~\mu$s half-life of $^{214}$Po makes it possible to suppress the background by using the coincident signals.

\subsection{Setup for Si-PIN radon detector}
\label{sec.pinsetup}
The Si-PIN Rn detector consists of a cylindrical electro-polished stainless steel vessel, a cylindrical high purity oxygen-free copper vessel and a Si-PIN photodiode. The dimensions of the copper vessel are 38~cm in diameter and 27.3~cm in height. A positive high voltage is supplied to the copper vessel to generate the electric field so that the positive daughters of $^{222}$Rn can be collected on the surface of the Si-PIN, then the decayed $\alpha$s can be detected by it~\cite{NIMA_421}. The Si-PIN photodiode is a Hamamatsu S3204-09 with a dimension of 18~mm $\times$ 18~mm, whose window has been removed to reduce energy loss as ¦Á particles strike the photodiode~\cite{s3204}. The 142A preamplifier~\cite{142A} made by ORTEC company is used for bias supply as well as for signal readout. Before sent into the oscilloscope(LeCroy 610Zi, 250MHz sampling frequency, 20${\mu}s$ readout window~\cite{lecroy610zi}) for pulse recording, the signal is amplified 50 times with an amplifier(ORTEC 671~\cite{671}).

\subsection{Setup for LS radon detector}
The LS Rn detector consists of an acrylic cylinder and viewed by two PMTs. The acrylic cylinder, with a dimension of 78~mm in diameter and 20~mm in height, is filled with liquid scintillator. The PMTs, 9821QB~\cite{pmt} from Electron Tubes company, directly face the two sides of the acrylic vessel. The coincident signals of ${\beta}$-$^{214}$Bi and ${\alpha}$-$^{214}$Po are used for background suppressing as well as for Rn concentration calculation.

\begin{figure}[htb]
\centering
\includegraphics[width=10cm]{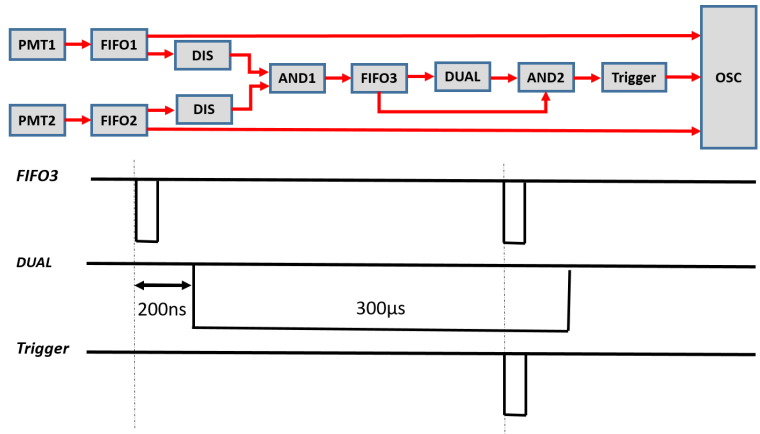}
\caption{Readout diagram of the LS Rn detector. }
\label{fig.electronics}
\end{figure}

\begin{figure}[htb]
\centering
\includegraphics[width=8cm]{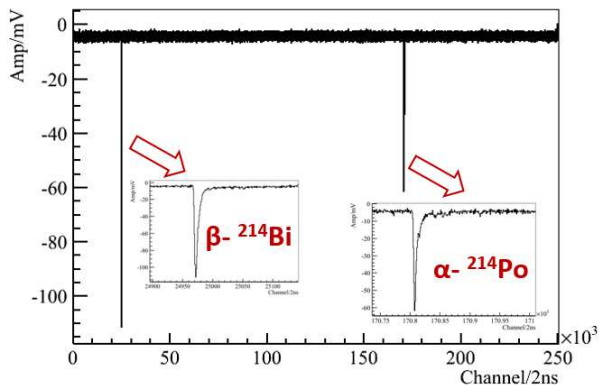}
\caption{An example signal of Rn in the LS detector. }
\label{fig.lssample}
\end{figure}

Figure~\ref{fig.electronics} shows the electronics scheme of the LS Rn detector. Signals from the two PMTs are divided into two paths by the quad linear Fan in/Fan out(FIFO, CAEN N625), one is sent into the oscilloscope(OSC) for pulse recording and the other one is used to form the trigger. The widths of the output pulses of the discriminator(DIS, CAEN N844) and the logic unit(AND, CAEN N455) are all 100~ns. The first pulse of AND1 is delayed by 200~ns and broadened to 300~${\mu}$s by the dual timer(DUAL, CAEN N93B). The coincidence of AND1 and DUAL is sent to the 500${\mu}$s time window oscilloscope as a trigger signal. In order to suppress the accidental coincident background, the single channel threshold is set to $\sim$80keV. Figure~\ref{fig.lssample} shows an example pulse of Rn signal.

\section{Optimization and calibration}
\label{sec.opt}
When $^{222}$Rn and its daughters are in equilibrium, the $^{222}$Rn concentration is proportional to the observed $\alpha$s or $\beta$s decayed from its daughters.   The event rate of $^{214}$Po is used to calculate the Rn concentration because there are no other $\alpha$ sources in its signal region~\cite{arxiv1704} as well as $^{214}$Po is known to have a higher collection efficiency than $^{218}$Po~\cite{NIMA_497}. While for LS Rn detector, the coincident signals of $\beta$-$^{214}$Bi and $\alpha$-$^{214}$Po are used for $^{222}$Rn concentration calculation.

In order to get the corresponding relationship between counting rate and $^{222}$Rn concentration, a gas flow solid Rn source, which is made by Source China University, is used to calibrate the detectors. When the gas flow through the gas flow solid Rn source with a flow rate of 1~liter/min, the Rn concentration in the output gas is 84.80$~\pm$~4.55~Bq/m$^3$, which is measured by RAD7~\cite{rad7}. The output gas can be used to calibrate the two kinds of detectors.
\subsection{Si-PIN Rn detector optimization and calibration}
\begin{figure}[htb]
\centering
\includegraphics[width=6cm]{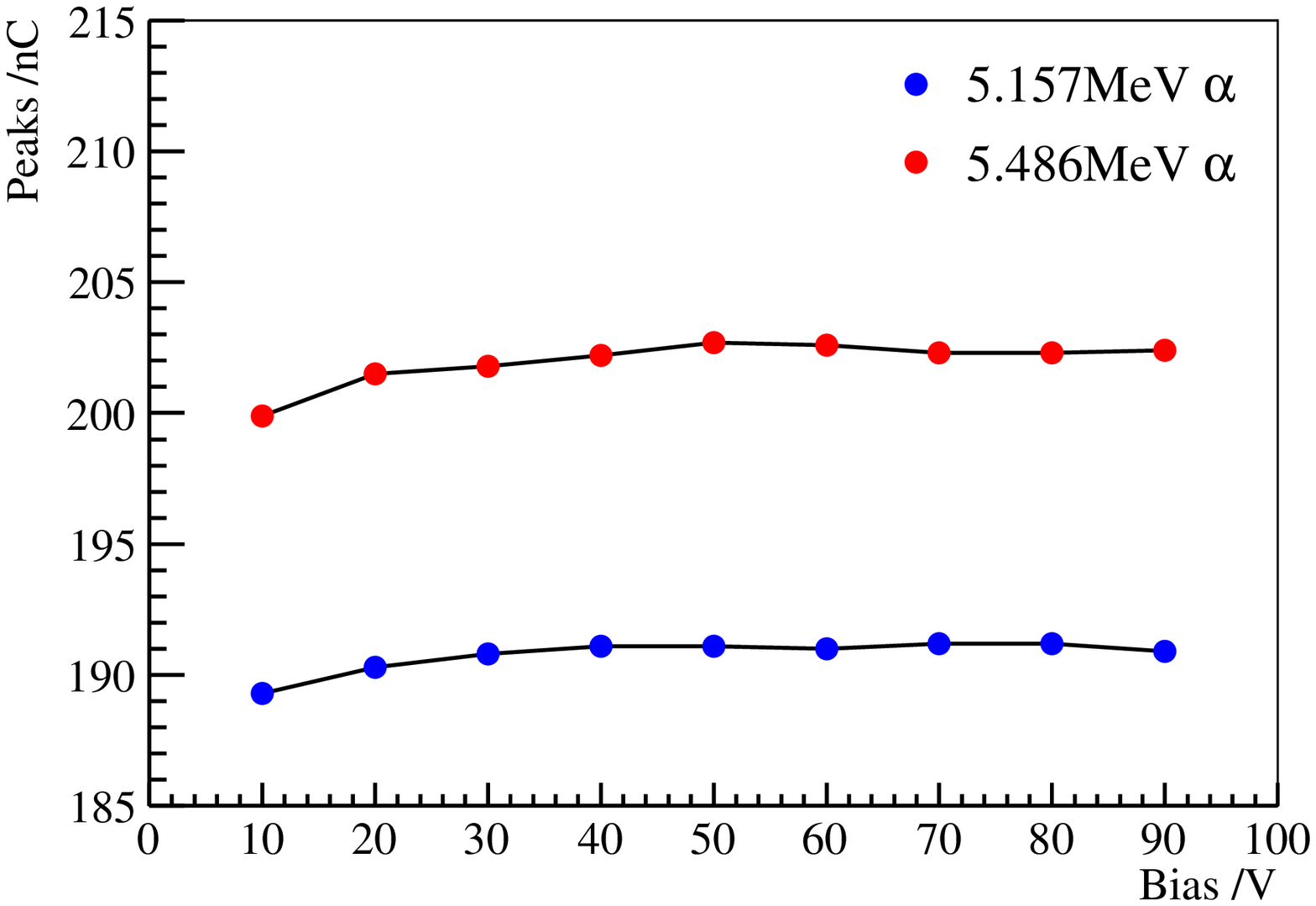}
\includegraphics[width=6cm]{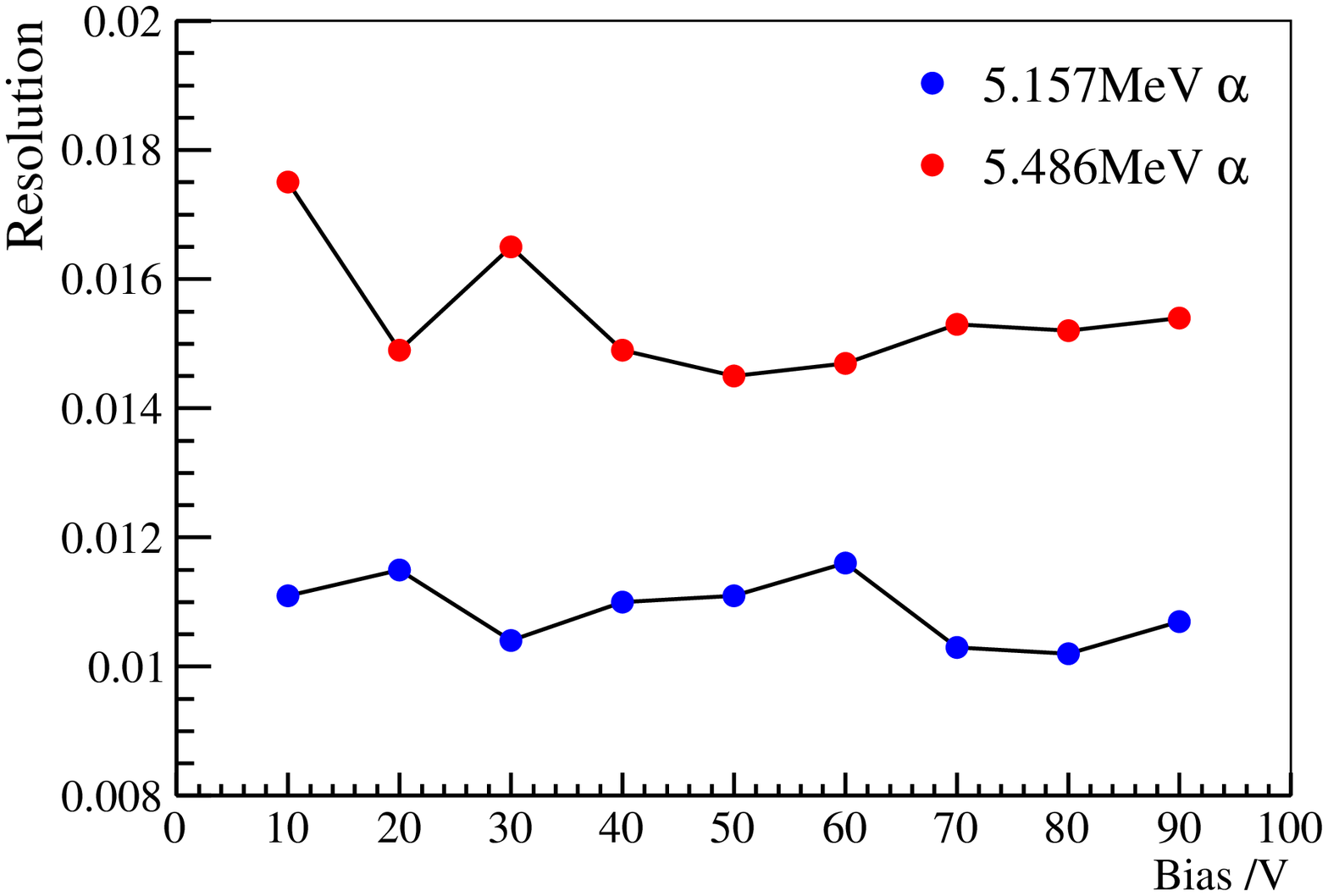}
\caption{Left: Si-PIN peak value VS bias. Right: Energy resolution VS bias.}
\label{fig.energy response}
\end{figure}

In order to select the working bias voltage for the Si-PIN photodiode, a $^{214}$Am/$^{239}$Pu hybrid $\alpha$ source is used to study its energy response. Figure~\ref{fig.energy response} shows the dependence of energy response and resolution with the applied bias voltage, which indicates that the Si-PIN photodiode has a stable performance at various bias voltage, thus based on the datasheet~\cite{s3204}, 60~V is chosen for other calibrations and measurements.

 A calibration factor($C_f$) is defined as Eq.~\ref{Eq.CF}:

\begin{equation}
C_f[(counts/h)/(Bq/m^3)]= \frac{measured ~^{214}Po~signal~rate}{^{222}Rn~concentration}
\label{Eq.CF}
\end{equation}

\begin{figure}[H]
\centering
\includegraphics[width=8cm]{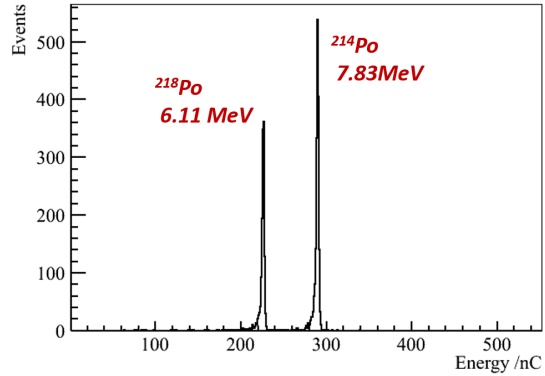}
\caption{Energy spectrum of Rn source. }
\label{fig.energyspectrum}
\end{figure}

Where the numerator is the measured $^{214}$Po signal rate on the Si-PIN photodiode in unit of counts/h, while the denominator is the $^{222}$Rn concentration in unit of Bq/m$^3$. The initial value of $C_f$, which is derived from the observed $^{214}$Po counting rate and the well-measured Rn gas, could be obtained from the Rn source calibration data. Figure~\ref{fig.energyspectrum} shows the energy spectrum of Rn source and the energy region of $\pm4\sigma$ around the $^{214}$Po peak is used to calculate the $C_f$.

\begin{figure}[H]
\centering
\includegraphics[width=8cm]{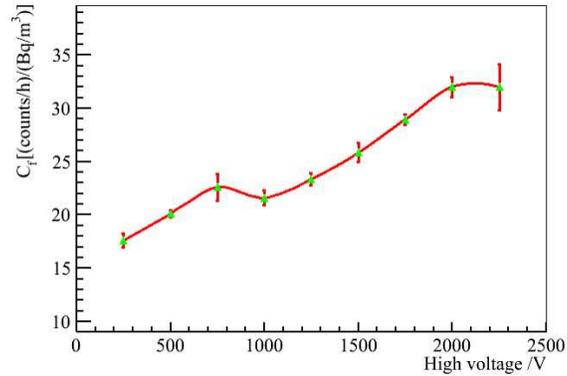}
\caption{$C_f$ dependence with supplied high voltage. }
\label{fig.highvoltage}
\end{figure}

As described in Sec.~\ref{sec.pinsetup}, a high voltage has to be applied on the oxygen-free copper vessel to generate the electrical field to collect the daughters of $^{222}$Rn, thus the high voltage value will affect the collection efficiency and the $C_f$ will vary from different high voltage values. In order to find the most suitable value, the relationship between $C_f$ and supplied high voltage has been measured, which is shown in Fig~\ref{fig.highvoltage}. As can be seen from the plot, the $C_f$ rises as the high voltage increases. Considering the stability of the detector and the collecting efficiency, 1500~V supplied high voltage is preferred and the $C_f$ is 26.38$\pm$1.56~(counts/h)/(Bq/m$^3$) for the background measurement.

\subsection{LS Rn detector calibration and MC simulation}

Liquid Scintillator has been used for Rn measurement for many decades because it has a strong enrichment capability for Rn as well as it is an excellent experimental target. As described in Sec.~\ref{sec.opt}, the counting rate of $\beta$~-~$^{214}$Bi and $\alpha$~-~$^{214}$Po coincident signals could be used to calculate the Rn concentration.

\begin{figure}[H]
\centering
\includegraphics[width=8cm]{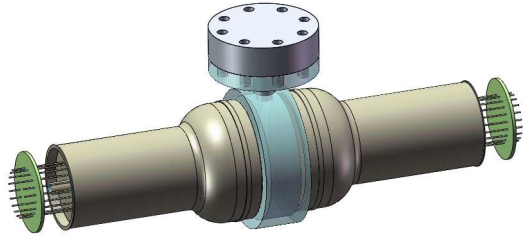}
\caption{The geometry constructed by the program. }
\label{fig.lsgeant4}
\end{figure}

\begin{figure}[htb]
\centering
\includegraphics[width=8cm]{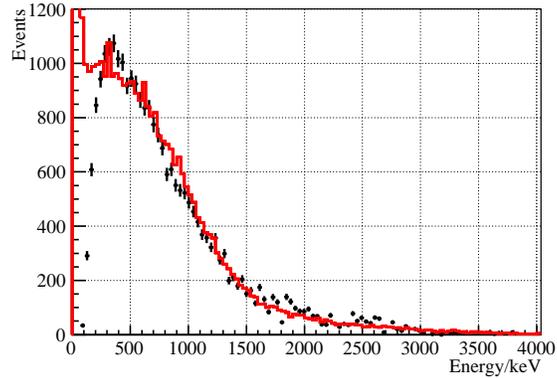}
\caption{The energy spectra of $\beta$-$^{214}$Bi. Red line: Simulated spectrum. Black dots with error bar: Measured spectrum.}
\label{fig.betaspectra}
\end{figure}

In order to study the energy deposit of $\beta$~-~$^{214}$Bi in LS, a Monte Carlo(MC) simulation with Geant4, which is based on the Daya Day simulation code, has been constructed. Figure~\ref{fig.lsgeant4} shows the geometry of the LS Rn measurement chamber. The effects of  LS vessel geometry, $^{214}$Bi distribution in LS as well as electronic response are included in the simulation. The simulated and measured $\beta$-$^{214}$Bi energy spectra are shown in Fig.~\ref{fig.betaspectra}, which are basically consistent with each other. The difference in the lower part is caused by the threshold and the detecting efficiency.

\begin{figure}[H]
\centering
\includegraphics[width=8cm]{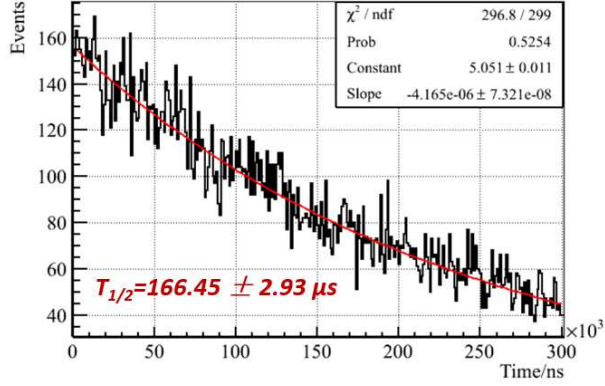}
\caption{The time difference between  $\beta$~-~$^{214}Bi$ and $\alpha$~-~$^{214}Po$. }
\label{fig.timedifference}
\end{figure}
In order to test the accuracy and efficiency of the electric trigger, which is shown in fig.~\ref{fig.electronics}, the time difference between $\beta$~-$^{214}$Bi and $\alpha$~-$^{214}$Po has been calculated and the distribution of the time difference as well as fitting result are shown in Fig~\ref{fig.timedifference}. The measured data are in good agreement with the theoretical expectation, which indicates that the LS system is a good Rn concentration measurement system.

For LS Rn detector, the $C_f$ is slightly different, the numerator in Eq.~\ref{Eq.CF} is changed into the counting rate of the coincident signals. By using the same Rn source with the Si-PIN Rn detector, the $C_f$ for LS Rn detector, which is 6.20$\pm$0.36~(counts/h)/(Bq/m$^3$), is derived from the calibration data.

\section{Sensitivity estimation and prospect}
The background measurement of Si-PIN Rn detector and LS Rn detectors are performed to estimate the sensitivity. The background event rate of $^{214}$Po for Si-PIN, which is derived from $87.4$ hours data, is 0.39$\pm$0.067~counts/h.  while for LS Rn detector, the background event rate of the coincident signals is 1.00$\pm$0.13~counts/h, which is derived from 55.0 hours data. At 90\% confidence level, the sensitivity can be estimated based on Eq.~\ref{Eq.sensitivity}~\cite{NIMA_867}:

\begin{equation}
L = \frac{1.64 * \sigma_{BG}}{C_f}
\label{Eq.sensitivity}
\end{equation}

where L is the sensitivity, $\sigma_{BG}$ is the uncertainty of the background event rate and $C_f$ is the calibration factor. Therefore, the sensitivity  is aournd 9.0~mBq/m$^3$ and 64.0~mBq/m$^3$ for a single day measurement for Si-PIN and LS Rn detectors respectively.

According to Ref~\cite{Rn_ratio_rad7}, when the water and the vapor are at diffusion equilibrium, the Rn concentration in the two media (water/vapor) is correlated by the Ostwald Coefficient and the ratio of radon concentrations in water to vapor follows Eq.~\ref{Eq.Rn_ratio}:

\begin{equation}
R = 0.105 + 0.405e^{-0.0502T}
\label{Eq.Rn_ratio}
\end{equation}

where R stands for the ratio and T is the temperature in unit of centigrade. For JUNO prototype, the temperature of the water has been kept at around 20 centigrade, thus R is around 0.25. Therefore, the Rn concentration in water can be derived from the gas measurement result.

Both of the Si-PIN Rn detector and the LS Rn detector are working well with the JUNO prototype at present and both of them have the potential to be developed as an online Rn concentration monitoring device for JUNO veto detector.

\section{Acknowledgements}
This work is supported by Xie Jialin Foundation of Institute of High Energy Physics(IHEP, Y7546150U2). Many thanks to Xiangcheng Meng, Guangpeng An of IHEP and Shoukang Qiu of University of South China for their help during the experiment.

\end{document}